# The Exit-Wave Power-Cepstrum Transform for Scanning Nanobeam Electron Diffraction: Robust Strain Mapping at Subnanometer Resolution and Subpicometer Precision

*Elliot Padgett[1], Megan E. Holtz[1,2], Paul Cueva[1], Yu-Tsun Shao[1], Eric Langenberg[2], Darrell G. Schlom[2,3], David A. Muller[1,3]*

[1] School of Applied and Engineering Physics, Cornell University, Ithaca, New York 14853, United States
[2] Department of Materials Science and Engineering, Cornell University, Ithaca, New York 14853, United States
[3] Kavli Institute at Cornell for Nanoscale Science, Ithaca, New York 14853, United States

*Abstract*

Scanning nanobeam electron diffraction (NBED) with fast pixelated detectors is a valuable technique for rapid, spatially resolved mapping of lattice structure over a wide range of length scales. However, intensity variations caused by dynamical diffraction and sample mistilts can hinder the measurement of diffracted disk centers as necessary for quantification. Robust data processing techniques are needed to provide accurate and precise measurements for complex samples and non-ideal conditions. Here we present an approach to address these challenges using a transform, called the exit wave power cepstrum (EWPC), inspired by cepstral analysis in audio signal processing. The EWPC transforms NBED patterns into real-space patterns with sharp peaks corresponding to inter-atomic spacings. We describe a simple analytical model for interpretation of these patterns that cleanly decouples lattice information from the intensity variations in NBED patterns caused by tilt and thickness. By tracking the inter-atomic spacing peaks in EWPC patterns, strain mapping is demonstrated for two practical applications: mapping of ferroelectric domains in epitaxially strained $PbTiO_3$ films and mapping of strain profiles in arbitrarily oriented core-shell Pt-Co nanoparticle fuel-cell catalysts. The EWPC transform enables lattice structure measurement at sub-pm precision and sub-nm resolution that is robust to small sample mistilts and random orientations.

## 1    Introduction

Characterization of nanoscale variations in lattice structure is essential for understanding the properties of many materials systems. (Scanning) transmission electron microscopy ((S)TEM) provides a powerful toolset for this, with several techniques for mapping local crystal structure, strain, and lattice distortions. [1–3] Aberration correctors have made atomic resolution STEM and TEM imaging widely available and effective for direct measurement of real-space lattice structure at the single-atom level. However, direct atomic resolution imaging is limited to relatively small fields of view (FOVs) around ~100 nm and requires a high beam dose and a specimen oriented on a high-symmetry zone axis. Furthermore, the quality of atomic resolution images is highly sensitive to instabilities such as sample drift, vibration, electromagnetic interference, and scan distortion.

Recent advances in fast, high-dynamic-range pixel-array detectors [4–6] are making scanning diffraction techniques, including nanobeam electron diffraction (NBED) and convergent beam electron diffraction (CBED), a promising approach for characterization of local lattice structure.[7] In NBED, a relatively low convergence angle is used such that the diffracted disks do not overlap and can be measured similarly to conventional TEM diffraction.[8] In CBED strain mapping, a highly convergent probe is used and the strain is extracted through measurement of the higher order Laue zone (HOLZ) lines formed by dynamical diffraction effects.[9] Both CBED and NBED techniques allow high precision, high spatial resolution lattice structure measurements over FOVs reaching up to micrometer length scales while imposing fewer constraints on stability.[10,11]

NBED is compatible with a wide range of sample thicknesses and orientations, making it a versatile scanning diffraction technique.[7,12] However, several challenges arise in applying NBED for structural characterization of complex specimens. High-spatial-resolution NBED measurements require a relatively high convergence angle, resulting in broad diffracted disks. In this case, intensity variations caused by dynamical diffraction and sample mistilts can hinder accurate and precise measurement of the disk centers. [13–15] Effective methods for extracting lattice information must ensure high computational efficiency, precision with limited beam dose, and robustness under nonideal conditions including background signals, varying peak intensities, and internal structure within the diffracted disks. One experimental technique that has been successfully deployed to improve NBED patterns is precession electron diffraction (PED), where the incident beam is tilted off of the optic axis of the microscope and precessed around the optic axis during NBED pattern acquisition.[16,17] PED can increase the number of higher order diffraction spots present by bringing the Ewald sphere to intersect more reciprocal lattice points, while also averaging out dynamical effects to create a more kinematic-like diffraction pattern with more internally uniform diffraction spots.[18–20] However, employing precession methods for NBED adds experimental complexity, increases acquisition time, and limits the achievable spatial resolution. Furthermore, precise localization of broad diffracted disks is a difficult computational task even under the relatively ideal conditions provided by PED. While a wide variety of techniques for data processing and disk localization have been employed – including correlation methods,[10,15] iterative fitting methods,[13,21] center of mass calculation,[11] and the circular Hough transform [22] – these challenges remain a limitation to the application of NBED to complex materials systems.

In this paper we propose a transformation to aid in the extraction of lattice structure and strain information from nanobeam diffraction maps, which can be applied both with and without PED. Our approach is inspired by the use of cepstral analysis in signal processing. Cepstral analysis was developed for echo detection and pitch determination, and was subsequently used for a wide variety of audio and image processing applications, especially the analysis of speech. [23–25] In audio processing, to track the time evolution of an audio signal, a short segment is selected using a time-domain window function and processed using short-time Fourier analysis. In particular, the power cepstrum (PC) of a time-domain function is calculated by first taking the logarithm of the power spectrum to produce a frequency-domain function. By then taking the power spectrum of this frequency-domain function, a new real-space function, similar to the autocorrelation function, is obtained; this is the PC.

The PC provides a very useful separation property for signals that are convolved in the time domain. In the frequency domain these signals become multiplicative, and then additive after the application of a logarithm. The originally convolved functions remain additive after the second Fourier transform to the time-domain PC, allowing convenient, independent measurement of signals with differing time-scales.

The measurement of crystal structure with NBED is analogous to short-time Fourier analysis of audio signals, with the goal instead being the determination of spatial variations in crystal structure. In NBED, a small real-space area is selected by the probe, and the diffraction pattern reveals its periodic structure in reciprocal space. The diffraction process physically "calculates" the Fourier transform of the real-space structure. Several of the problems complicating lattice structure measurements from NBED patterns, including intensity variations from sample mistilt and multiple scattering, can be approximated as multiplicative in diffraction space (and therefore convolutions in real space). Analogy to audio signal processing suggests that a real-space cepstrum may be useful for separating these effects from the lattice structure of interest. Because the first steps of producing a real-space cepstrum are performed physically by the diffraction process, a real-space cepstrum can be calculated by taking the power spectrum of the log-scaled NBED pattern. As shown in Section 2, this is equivalent to the power cepstrum of the exit wave, and so we denote this transform the exit wave power cepstrum (EWPC).

The process of calculating the EWPC is illustrated in **Figure 1**. **Figure 1**(a) shows a NBED pattern for a face-centered cubic Pt-Co alloy oriented on the [110] zone axis with the direct beam saturated in post-processing for display purposes. **Figure 1**(b) shows the logarithm of the NBED pattern, which flattens the intensity variation between the diffraction spots and brings more spots into the intermediate intensity range. **Figure 1**(c) shows the EWPC transform of the NBED, or the magnitude of the Fourier transform of **Figure 1**(b). The EWPC pattern is similar to an autocorrelation of the real lattice, i.e. a Patterson function, [26] and concentrates the signal from the diffracted disks into sharp real-space points corresponding to inter-atomic spacings. This transform makes precise fitting of the peaks for measurement of the lattice structure numerically convenient and separates the effects of tilt and thickness on the NBED pattern from the lattice structure information.

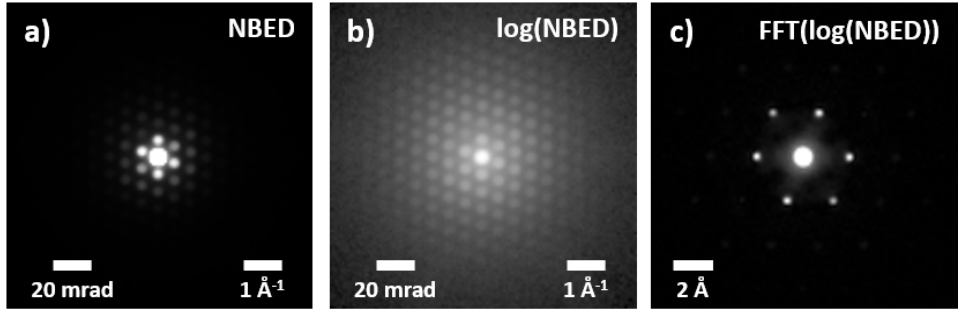

**Figure 1:** Illustration of the EWPC transform for an FCC Pt-Co alloy oriented on the [110]-zone axis. (a,b) Nanobeam electron diffraction (NBED) pattern shown with (a) linear and (b) logarithmic intensity scales. (c) The EWPC transform for this diffraction pattern, calculated as the fast Fourier transform (FFT) of the log of the NBED pattern.

In this manuscript we present a simple model of the EWPC transform in the context of NBED and experimentally demonstrate its key features for strain mapping. Using Bloch wave simulations to incorporate dynamical diffraction effects, we investigate the robustness of the approach to varying sample thickness and mistilt. We then demonstrate its practical value for two applications: ferroelectric domains in epitaxially strained $PbTiO_3$ films and core-shell Pt-Co nanoparticle catalysts. For the $PbTiO_3$ films we show EWPC mapping is robust to large mistilts caused by ferroelastic strain. For the Pt-Co nanoparticles we show that the EWPC transform enables sub-nm resolution strain mapping that is robust to random particle orientations and suitable to provide strain measurements in ~nm thick shells. These examples show that the EWPC transform is a powerful tool for high spatial resolution NBED strain mapping that is precise, dose efficient, and robust to a variety of sample complexities, suppressing artefacts from thickness fringes and bend contours.

## 2   Theory of the EWPC

### 2.1   Cepstral Analysis and the Exit Wave Power Cepstrum

Cepstral analysis was originally developed for the analysis of human speech, with the goal of determining the time evolution of the pitch in audio signals.[23] Cepstral analysis provides a means for separating the fundamental pitch frequency from convolved formants that shape the wave. In particular, the power cepstrum of a function $f(t)$ is defined as the power spectrum of the log of the power spectrum of $f(t)$:

$$\text{PC}(f(t)) = \left| \mathcal{F}\left( \ln\left( |\mathcal{F}(f(t))|^2 \right) \right) \right|^2, \tag{1}$$

where $\mathcal{F}$ denotes a Fourier transform. For convolved functions $\alpha(t) * \beta(t)$ with Fourier transforms $A(\omega) = \mathcal{F}(\alpha(t))$ and $B(\omega) = \mathcal{F}(\beta(t))$, the power cepstrum provides a useful separation property:

$$\begin{aligned}\text{PC}(\alpha(t) * \beta(t)) &= \left| \mathcal{F}(\ln(|A(\omega) \cdot B(\omega)|^2)) \right|^2 \\ &= \left| \mathcal{F}(\ln(|A(\omega)|^2) + \ln(|B(\omega)|^2)) \right|^2\end{aligned}$$

$$= \left| \mathcal{F}\left(\ln\left(|\mathcal{F}(\alpha(t))|^2\right)\right) + \mathcal{F}\left(\ln\left(|\mathcal{F}(\beta(t))|^2\right)\right) \right|^2. \tag{2}$$

For $\alpha(t), \beta(t)$ with well-separated frequency bands, this allows additive separation of the convolved signals,

$$\text{PC}(\alpha(t) * \beta(t)) \approx \text{PC}(\alpha(t)) + \text{PC}(\beta(t)), \tag{3}$$

as desired for independent analysis of pitch and formants.

This situation is closely analogous to the problem of lattice mapping using nanobeam diffraction, where our goal is to measure the spatial variation of the periodic lattice structure in a sample. Sample tilt and thickness create modulations in the intensity of diffracted beams, similar to the effect of formants on an audio power spectrum, which we would ideally separate from the lattice information. These similarities suggest that some form of cepstral analysis may be useful in nanobeam diffraction mapping as well.

To determine how to structure the analysis, it is useful to consider the trade-off presented by the choice of probe size in nanobeam diffraction. A probe that provides high localization in real space will result in delocalization of the reciprocal lattice information in diffraction space by broadening diffracted peaks into disks. As a result, measurement of the lattice structure in diffraction space may become imprecise or computationally inconvenient. A transform into real space can avoid these challenges for lattice measurement.

To take advantage of the cepstral separation properties and real space lattice measurement, we will define the exit wave power cepstrum (EWPC) as follows:

$$\text{EWPC}_\phi(x) = \text{PC}(\phi(x)) = \left| \mathcal{F}\left(\ln\left(|\mathcal{F}(\phi(x))|^2\right)\right) \right|^2 = \left| \mathcal{F}\left(\ln(I(q))\right) \right|^2, \tag{4}$$

for exit wave $\phi(x)$ with real-space position $x$ and reciprocal-space scattering vector $q$. The EWPC can easily be calculated from the diffraction pattern $I(q) = |\mathcal{F}(\phi(x))|^2$, which "calculates" the first part of the power cepstrum transform physically. The EWPC concentrates the signal from the set of periodic diffracted disks into sharp real-space points that are convenient for high-precision lattice measurements while also separating out "undesirable" effects from specimen tilt and thickness, as we will discuss in detail below.

## 2.2 Interpretation of EWPC Patterns

We will use a simple model of nanobeam electron diffraction to qualitatively illustrate the separation properties of the EWPC and provide a basis for the interpretation of EWPC patterns for diffraction mapping. Dynamical effects are considered using Bloch wave simulations in Section 4.2, and similar trends hold. To provide an illustrative and simplified analytical model of the real-space exit wave $\phi(x)$ and the EWPC we can begin by approximating the specimen as a strong phase object, such that:

$$\phi(x) = \phi_P(x)\exp(i\sigma\bar{V}(x)), \tag{5}$$

where $\phi_P(x)$ is the probe function, $\sigma$ is a scaling factor, and $\bar{V}(x)$ is the projected specimen potential. Foreshortening from geometric projection [29] is included in $\bar{V}$. The diffraction pattern is

$$I(q) = |\mathcal{F}(\phi(x))|^2$$
$$= |\mathcal{F}(\phi_P(x)) * \mathcal{F}(\exp(i\sigma\bar{V}(x)))|^2$$
$$= |\Phi_P(q) * \mathcal{V}(q)|^2, \quad (6)$$

where $\Phi_P(q)$ is the diffraction-space probe and $\mathcal{V}(q) = \mathcal{F}(\exp(i\sigma\bar{V}(x)))$. In the kinematical or weak phase approximation, we take $\exp(i\sigma\bar{V}(x)) \approx 1 + i\sigma\bar{V}(x)$, so that $\mathcal{V}(q)$ is essentially the Fourier transform of the projected specimen potential. In the strong phase approximation, the additional higher order terms can be thought of as multiple scattering corrections.

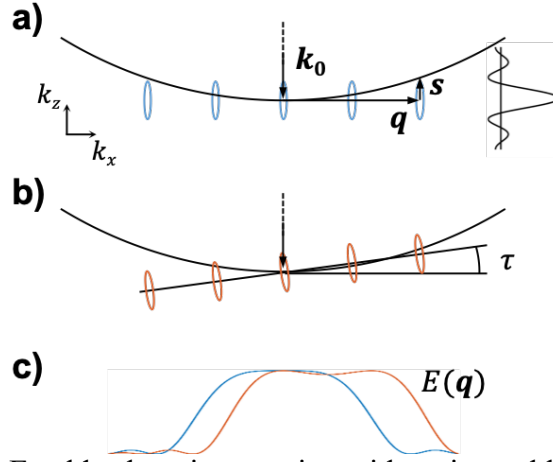

**Figure 2:** Diagram of the Ewald sphere intersecting with reciprocal lattice points broadened by finite specimen thickness for (a) an on-axis specimen and (b) a specimen tilted at angle $\tau$, with incident wavevector $k_0$, scattering vector $q$, and excitation error $s$. The curve at the right of (a) illustrates the reciprocal lattice point broadening. The coordinates are oriented with $k_x$ to the right, $k_y$ into the page, and $k_z$ up. (c) The resulting amplitude-attenuating envelope function $E(q)$ for each case.

The strong phase model can be extended to account for kinematical specimen tilt and thickness effects by inclusion of the Ewald sphere's intersection with the reciprocal lattice.[27,28] The reciprocal space distance between the Ewald sphere and the reciprocal lattice points is expressed by the excitation error $s(q)$, illustrated in Figure 2(a,b). For electrons with incident wavevector $k_0$, the excitation error for an on-axis specimen is $s(q) \approx q^2/2k_0$, where $q^2 = q_x^2 + q_y^2$. A tilt in the reciprocal lattice by an angle $\tau$ about the $k_y$-axis will add an additional term $\Delta s = -q_x \sin\tau \approx -q_x\tau$, for a total excitation error $s(q) \approx q^2/2k_0 - q_x\tau$.

A sample of thickness $T$ can be represented in real space by a top-hat function in the z-direction multiplied on the infinite lattice. In reciprocal space, this finite thickness broadens the reciprocal lattice points by convolving with the Fourier transform of the top-hat function, $\text{sinc}(k_z T/2)$. The excitation error thus leads to an attenuation of the diffracted amplitude[1]:

$$I(q) = |\Phi_P(q) * (E(q) \cdot \mathcal{V}_0(q))|^2, \quad (7)$$

---

[1] Here we have neglected phase shifts on the diffracted beams, which do not impact the diffracted intensities.

where $\mathcal{V}_0$ is the "tilt-free" object function, defined such that $\mathcal{V}(\boldsymbol{q}) = E(\boldsymbol{q}) \cdot \mathcal{V}_0(\boldsymbol{q})$, and

$$E(\boldsymbol{q}) = \text{sinc}\left(\frac{s(\boldsymbol{q})T}{2}\right) = \text{sinc}\left(\left(\frac{q^2}{2k_0} - q_x\tau\right)\frac{T}{2}\right). \tag{8}$$

The function $E(\boldsymbol{q})$, illustrated in Figure 2(c), can be viewed as an envelope function attenuating the diffraction pattern. For samples of finite thickness, multiple scattering will also tend to broaden the envelope of the diffraction pattern, although this effect is difficult to model analytically. This effect can also be included within $E(\boldsymbol{q})$ for purposes of this simple model. Typically, this envelope function varies slowly in comparison to the convergence angle of the probe[2], and so we can make the approximation:

$$I(\boldsymbol{q}) \approx \left|E(\boldsymbol{q}) \cdot \left(\Phi_P(\boldsymbol{q}) * \mathcal{V}_0(\boldsymbol{q})\right)\right|^2 = |E(\boldsymbol{q})|^2 \cdot |\Phi_P(\boldsymbol{q}) * \mathcal{V}_0(\boldsymbol{q})|^2. \tag{9}$$

This approximation allows for separation of the tilt envelope in the EWPC:

$$\begin{aligned}\text{EWPC}_\phi(\boldsymbol{x}) &\approx \left|\mathcal{F}\left(\ln(|E(\boldsymbol{q})|^2 \cdot |\Phi_P(\boldsymbol{q}) * \mathcal{V}_0(\boldsymbol{q})|^2)\right)\right|^2 \\ &= \left|\mathcal{F}\left(\ln(|E(\boldsymbol{q})|^2) + \ln(|\Phi_P(\boldsymbol{q}) * \mathcal{V}_0(\boldsymbol{q})|^2)\right)\right|^2.\end{aligned} \tag{10}$$

The second term above can be further simplified with the assumption of nanobeam diffraction conditions. In particular, we will assume that the probe $\Phi_P(\boldsymbol{q})$ is a flat, top-hat function in diffraction space, and the separation between Bragg points in $\mathcal{V}_0(\boldsymbol{q})$ is larger than twice the convergence semi-angle, so that the diffracted disks do not overlap. Under these conditions, the second term in Equation 10 is equivalent to a simpler form with the probe function outside of the logarithm:

$$\ln(|\Phi_P(\boldsymbol{q}) * \mathcal{V}_0(\boldsymbol{q})|^2) \approx \Phi_P(\boldsymbol{q}) * \ln(|\mathcal{V}_0(\boldsymbol{q})|^2). \tag{11}$$

With these simplifications, the EWPC becomes:

$$\begin{aligned}\text{EWPC}_\phi(\boldsymbol{x}) &\approx \left|\mathcal{F}\left(\ln(|E(\boldsymbol{q})|^2) + \Phi_P(\boldsymbol{q}) * \ln(|\mathcal{V}_0(\boldsymbol{q})|^2)\right)\right|^2 \\ &= \left|\mathcal{F}\left(\ln(|E(\boldsymbol{q})|^2)\right) + \mathcal{F}(\Phi_P(\boldsymbol{q})) \cdot \mathcal{F}\left(\ln(|\mathcal{V}_0(\boldsymbol{q})|^2)\right)\right|^2 \\ &= \left|\mathcal{F}\left(\ln\left(|\mathcal{F}(\epsilon(\boldsymbol{x}))|^2\right)\right) + \phi_P(\boldsymbol{x}) \cdot \mathcal{F}\left(\ln\left(|\mathcal{F}(v_0(\boldsymbol{x}))|^2\right)\right)\right|^2,\end{aligned} \tag{12}$$

where we have defined the real-space tilt-blur function $\epsilon(\boldsymbol{x}) = \mathcal{F}^{-1}(E(\boldsymbol{q}))$, and the real-space tilt-free object function $v_0(\boldsymbol{x}) = \mathcal{F}^{-1}(\mathcal{V}_0(\boldsymbol{q}))$.

We will generally have $\epsilon(\boldsymbol{x})$ and $v_0(\boldsymbol{x})$ with well-separated frequency bands, allowing us to assume no overlap between these terms:

$$\text{EWPC}_\phi(\boldsymbol{x}) \approx \text{PC}(\epsilon(\boldsymbol{x})) + |\phi_P(\boldsymbol{x})|^2 \cdot \text{PC}(v_0(\boldsymbol{x})). \tag{13}$$

This provides a simple interpretation of EWPC patterns. The lattice information in contained in the power cepstrum of the tilt-free real-space lattice $\text{PC}(v_0(\boldsymbol{x}))$. The lattice power cepstrum is similar to a Patterson function [26] or an autocorrelation of the real lattice, which also contains peaks corresponding to inter-atomic spacings, although by separating out the envelope function the EWPC provides sharper, more easily measured peaks. This difference is discussed further in the Appendix. Like a Patterson function, the lattice power cepstrum is higher symmetry than the real lattice, losing any inversion asymmetry. These peaks will be attenuated by a probe function envelope $|\phi_P(x)|^2$. This is intuitively appealing – we observe the interatomic spacings and orientations in the region illuminated by the probe – and sets a lower bound for the size of probe that can be used. The effect of the Ewald sphere intersecting the reciprocal lattice, which is sensitive to both sample tilt and thickness, is additively separated into the term $\text{PC}(\epsilon(x))$ and concentrated into the short-length region of the EWPC pattern.

---

[2] For a sample tilted far from a major zone axis, the assumption of a slowly varying envelope function degrades, and the effect of tilt will remain convolved into the EWPC to some degree rather than separating linearly, causing some blurring of the pattern.

For simplicity, the calculations of the EWPC in this paper will generally omit the outermost power of two from the definition. This exponent is sometimes included to give the power cepstrum a symmetric form to the power spectrum, where the power of two allows physical interpretation as an energy density per unit frequency for many applications. However, the power cepstrum does not carry any equivalent physical meaning, as it does not represent an energy distribution. For purposes of strain mapping this exponent has no practical importance because it does not alter the position of peak maxima, and simply acts as a gamma adjustment on EWPC patterns that can make secondary features more difficult to discern. All of our patterns are thus calculated from nanobeam diffraction patterns using the transform:

$$\text{EWPC} = \left|\mathcal{F}\left(\ln(I(\boldsymbol{q}))\right)\right|. \tag{14}$$

In practice, it is also useful to multiply the log-scale diffraction pattern by a window function to prevent the introduction of artifacts caused by non-periodic boundaries in diffraction space. A Hann window is used for this purpose in calculations throughout the paper.

### 2.3 Lattice and Strain Mapping Using EWPC Patterns

Lattice mapping can be achieved by calculating the positions of the inter-atomic spacing peaks in EWPC patterns. These peaks will include full-unit-cell translations, which will generally have the brightest peaks, as well as sub-unit-cell spacings. If peaks are identified to correspond to known lattice spacings, this provides a direct measurement of the local lattice parameters.

Systems with spatially varying lattice distortions or strain are of particular interest for high resolution lattice mapping. Calculation of lattice distortions or strain requires the selection of a reference structure. It is often ideal to use the local lattice parameters of the material in its undeformed state as a reference, and strain calculated using this convention is known as the "material strain". However, the local, undeformed lattice parameters are often not known, making this approach impractical. An alternative is to use a single fixed reference, giving a quantity known as the "Lagrange strain".[2] The Lagrange strain can be calculated from a lattice map alone, without the need for any prior knowledge of the crystal structure.

To determine the projected strain tensor, a 2X2 matrix $D$ is calculated that transforms the reference lattice points $\boldsymbol{p}$ onto the measured points $\boldsymbol{p}'$, i.e. $\boldsymbol{p}' = D \cdot \boldsymbol{p}$. This distortion matrix $D$ includes both rigid rotation of the lattice and strain deformation. These may be separated by polar decomposition:

$$D = R \cdot U = V \cdot R, \tag{15}$$

where $R$ is a unitary rotation matrix, and $U$ and $V$ are the non-rotational deformation transformations. The choice of $U$ or $V$ determines the coordinate system of the strain tensor: $U$ is in "lattice coordinates", corresponding to the local lattice directions, and $V$ in "image coordinates", corresponding to the fixed $q_x$ and $q_y$ directions in the diffraction pattern (independent of the lattice rotation). The strain tensor $\varepsilon$ is then calculated as:

$$\varepsilon = U - 1, \text{or } \varepsilon = V - 1. \qquad (16)$$

The components of the strain tensor, $\varepsilon_{xx}$, $\varepsilon_{yy}$, and $\varepsilon_{xy} = \varepsilon_{yx}$ are the $x$-direction stretch, $y$-direction stretch, and diagonal $xy$-direction stretch or shear, respectively. These quantities, along with the in-plane lattice rotation $\theta$ determined from $R$, are often of interest and can be examined directly. For some systems, such as core-shell nanoparticles, the geometry leads to varying strain directions that are not easily interpreted in rectangular coordinates. An alternative formulation is the "strain ellipse", which describes the strain through the semi-major axis, semi-minor axis, and angle of the ellipse formed by applying the strain transformation to a circle. This is mathematically equivalent to describing the strain in its local eigenbasis, with the eigenvalues defining the semi-major and semi-minor axes of the strain ellipse and the direction of the eigenvectors defining the angle of principal strain.

## 3 Materials and Methods

### 3.1 TEM Sample Preparation

Cross-sectional specimens of $PbTiO_3$ thin films on $DyScO_3$ and $GdScO_3$ substrates were prepared using focused ion beam lift-out and thinning with an FEI Strata 400 Focused Ion Beam with a final 2 keV milling step to reduce surface damage. Specimens of carbon-supported Pt-Co nanoparticle fuel cell catalysts were prepared by cross-sectioning a fuel cell membrane electrode assembly that had undergone an accelerated stability test. Details of catalyst, cell assembly, and stability testing are reported in another manuscript. [30] Cross-sections were cut using a Leica Ultracut UCT Ultramicrotome at a nominal thickness of 40 nm and collected on lacey carbon-coated TEM grids. The monolayer $MoS_2$ specimen was suspended on a holey amorphous silica support film.

### 3.2 Scanning Transmission Electron Microscopy (STEM)

STEM imaging, electron energy loss spectroscopy (EELS), and diffraction mapping were performed in a FEI Titan Themis (S)TEM operated at 300 kV (120 kV for 2D materials). Atomic resolution imaging was performed using a convergence semi-angle of 21.4 mrad with the image signal collected on a high angle annular dark field (HAADF) detector. EELS mapping of Pt-Co nanoparticles was performed using a Gatan GIF Quantum 965 spectrometer in single-range EELS mode. Maps were acquired with a convergence semi-angle of 21.4 mrad, a ~200 pA beam current, a 10 ms dwell time, and ~1 Å pixel size, leading to a total dose of ~$10^9$ e$^-$/nm$^2$. Composition maps were extracted by integrating the signal from the Co $L_{2,3}$ and Pt $M_{4,5}$ edges after background subtraction using exponential and linear-combination-of-power-laws background fits, respectively. EELS calculations were performed in MATLAB using functions in the image processing toolbox.

Lattice maps from atomic resolution STEM imaging were calculated using high-quality images made from 12-24 cross correlated acquisitions taken with a short dwell time (0.5-2 us) to minimize drift and improve signal. Using a 40 pA beam, this corresponded to a typical beam dose around $10^7$ e$^-$/nm$^2$. Rough atomic positions were found by segmenting the image using a threshold after morphological background subtraction. Then, precise atomic positions were found by fitting with a two-dimensional Gaussian function near each rough atomic position. The atomic columns for Pb and Ti were then separated based on their HAADF intensity. Strain calculations were made from the Pb atomic positions following the same conventions detailed in the following section.

## 3.3 STEM Nanobeam Diffraction Mapping

Nanobeam diffraction maps were acquired in STEM microprobe mode using the EMPAD pixelated detector, which has high dynamic range and single electron sensitivity to allow collection of all scattered electrons as well as the unsaturated direct beam. [4] Typical acquisitions used a 1 ms exposure time and 256X256 real-space pixels. With the ~1 ms detector readout time this results in a total acquisition time of ~2 minutes. Typical doses ranged from $10^3$ – $10^6$ e$^-$/nm$^2$ depending on the required resolution and field of view. The detector reads out 128X128 pixels per frame, (with 124X124 used for diffraction pattern acquisition,) in 32-bit floating point numbers, resulting in ~4 GB files for typical acquisitions.

## 3.4 Calculation of Nanobeam Diffraction Lattice Maps

All computation was performed in Matlab. Essential functions and example codes for power-cepstral STEM (PC-STEM) analysis are available on Github at https://github.com/muller-group-cornell/PC-STEM. Prior to calculating the EWPC transform, a constant value is added to the raw diffraction patterns to make all values greater than zero for numerical convenience when computing logarithms. For calculation of 2D EWPC transforms a discrete fast Fourier transform is used, with a 2D Hann window to minimize edge artifacts.

Peaks are identified for lattice mapping using small rectangular masks that include the peak maximum for all pixels in the region of interest. At least two peaks must be tracked. At each real space pixel and for each peak mask, the mapping routine first determines a coarse (1-pixel precision) peak position from the pixel with maximum EWPC intensity within the mask. A high precision (sub-pixel) measurement of the peak maximum is then calculated using an optimization routine initialized at the coarse peak position. The objective function for the optimization calculates the value of the EWPC transform at a single point using a continuous Fourier transform, as well as the same constant offset and 2D Hann window used for 2D EWPC calculations. Optimization was performed using the Nelder-Mead simplex method (Matlab's fminsearch). The optimization algorithm is constrained within the peak region using a modified objective function. The convergence tolerance of the optimization algorithm can also be adjusted to improve computation time. Strain mapping calculations were performed on a high-end desktop computer (2015 iMac 27"; 4.0GHz Quad-core Intel Core i7; 32GB 1600MHz DDR3 SDRAM). Typically ~10 ms of computation time was required per EWPC peak fit, allowing strain map calculations within a few minutes.

After peak maxima are located for the entire map, lattice parameters may be calculated directly, or lattice distortions may be tracked by calculation of a 2X2 strain tensor. For calculation of the strain tensor, a transformation matrix $D$ is first determined for each real-space pixel by fitting the peak positions to reference points, (using Matlab's fitgeotrans,) either absolute lattice vectors for known structures or relative, internal references for the Lagrange strain.

### 3.5 Growth of PbTiO$_3$ films

High crystal quality PbTiO$_3$ films were grown on 110-oriented DyScO$_3$ and GdScO$_3$ substrates by reactive molecular beam epitaxy (MBE) in a Veeco GEN 10 system using distilled ozone ($\approx$80% mol O$_3$) as an oxidant and elemental Pb and Ti as source materials, with typical fluxes of $1.7 \times 10^{14}$ and $1.5 \times 10^{13}$ atoms/(s·cm$^2$). Distilled ozone pressure in the MBE chamber and the substrate temperature were fixed to $9 \times 10^{-6}$ Torr and 600°C, respectively. Pb and Ti were continuously codeposited, achieving the alternation of PbO and TiO$_2$ atomic layers by adsorption-control. The layer-by-layer growth and the adsorption control growth window is continuously monitored by reflection high energy electron diffraction (RHEED). After deposition, the films were cooled down at $\approx$100°C/min, while setting the distilled ozone pressure at $5 \times 10^{-6}$ Torr.

### 3.6 Diffraction Simulations

The diffraction simulation is based on the Bloch-wave method [2] with 1.3 mrad semi-convergence angle, using the neutral atomic scattering factors of Doyle & Turner [31], absorption parameters of Bird & King[32], a total of 101 beam and 797 incident beam directions within the diffracted disc. For testing the EWPC at different diffraction conditions, the crystal is tilted in 0.05° increments from exact incidence along the [001] zone axis up to 3°. At each tilt, diffraction patterns were simulated for different sample thicknesses.

## 4 Results and Discussion

### 4.1 Properties of EWPC Patterns

A series of experiments was conducted to validate and illustrate the model presented in the theory section and to guide use of EWPC patterns for strain mapping. The EWPC transform leverages the additive separation properties provided by the log-scaled NBED to separate the effects of tilt and thickness from useful lattice information. **Figure 3** illustrates these properties for NBED patterns (a) taken from a$_1$/a$_2$ domains in a PbTiO$_3$ film that are oriented on the [100] zone axis (left) and tilted ~60 mrad off axis (right). As the EWPC transform (**Figure 3**b) is the Fourier transform of the log-scaled NBED, it can be used for filtering as done in conventional Fourier filtering. The slowly-varying envelope function $E(\boldsymbol{q})$ formed by tilt and thickness effects has its information concentrated into the short-distance region near the origin in the EWPC space. Selecting this region (highlighted in red in **Figure 3**b) with a low-pass filter of the log-scaled NBED isolates the envelope, as shown in **Figure 3**c. The lattice information is concentrated in the peaks at longer distances in the EWPC space. Selecting these with a high-pass filter of the log-scaled NBED isolates the periodic diffraction spots, as shown in **Figure 3**d.

These two components can be additively recombined to reconstruct the original log-scaled NBED. The relative tilt between the left and right halves of **Figure 3** primarily changes the envelope function, impacting short-distance region of EWPC separately from lattice information. However, the separation is imperfect for such a high tilt angle because the assumption of a slowly-varying envelope function required for complete separation is violated. This is visible in the slight blurring of the EWPC peaks at the right in **Figure 3**b, although these peaks remain sharp and suitable for fitting to measure the lattice structure.

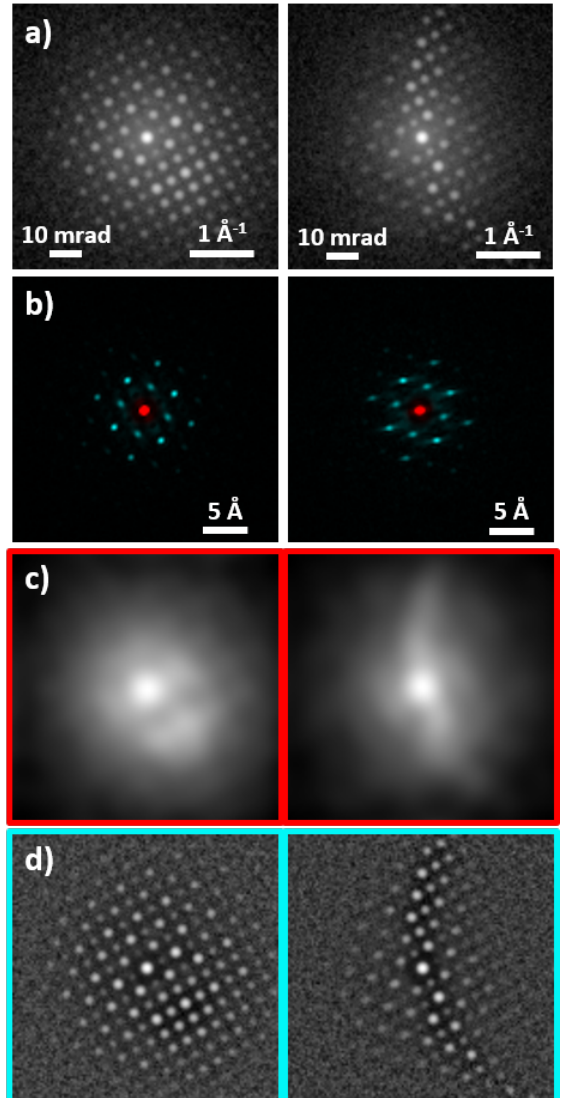

**Figure 3:** Illustration of additive separation of the tilt-envelope and the lattice information using the EWPC transform from a $PbTiO_3$ film on $GdScO_3$. (a) Log-scaled NBED patterns for $PbTiO_3$ oriented on [100] axis (left) and tilted ~60 mrad off axis. (b) EWPC patterns corresponding to the NBED patterns in (a). The short-distance region is selected with a low-pass filter shown by red coloring in (b), and the resulting low-pass filtered log(NBED) pattern (c) corresponds to the envelope function. The long-distance region is selected with a high-pass filter shown by blue coloring in (b), and the resulting high-pass filtered log(NBED) pattern (d) corresponds to the periodic diffraction information. The patterns in (c) and (d) additively recombine to the log(NBED) pattern in (a).

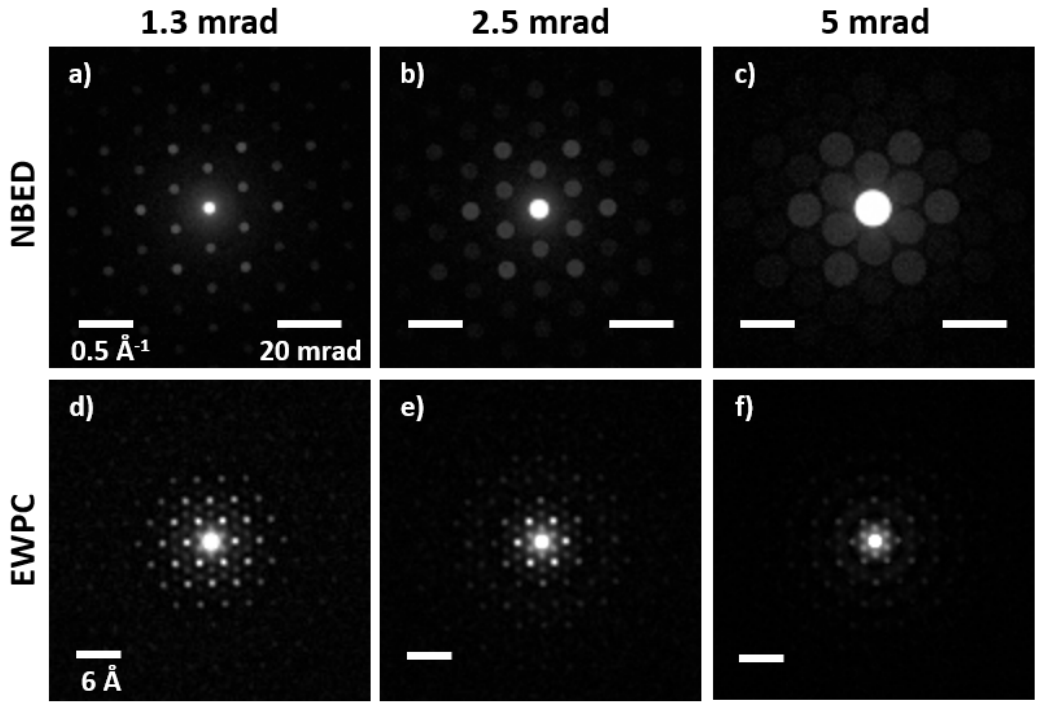

**Figure 4:** Diffraction patterns (a-c) and corresponding EWPC patterns (d-f) for different convergence semi-angles of the STEM probe acquired with 120 keV beam energy on monolayer $MoS_2$. Convergence semi-angles are (a,d) 1.3 mrad, (b,e) 2.5 mrad, (c,f) 5 mrad, corresponding to diffraction-limited resolutions (FWHM probe diameter) of 1.6 nm, 0.8 nm, and 0.4 nm, respectively.

The choice of STEM convergence angle in NBED determines both the real-space resolution and the size of the diffracted disks, which may impact the precision and ease of calculation of lattice structure. **Figure 4** illustrates the impact of the convergence angle on EWPC patterns (d-f) calculated from NBED patterns (a-c) from monolayer $MoS_2$. As shown in Equation 13, the probe creates an envelope function multiplied over the lattice structure peaks in the EWPC pattern. Smaller convergence angles give a wider probe that makes more inter-atomic spacing peaks apparent in the EWPC pattern. Larger convergence angles give a smaller probe, attenuating the intensity of higher-order spacings. In contrast to direct diffraction, the intensity of the peaks, and not their localization, is affected by the convergence angle. Measuring more higher-order peaks in a lattice structure may improve precision of EWPC analysis. It is notable that weak higher order peaks are also present in the tails of the probe envelope, although these peaks represent lattice structure information that is more delocalized than the lower order peaks within the central part of the probe envelope.

The upper bound on the spatial resolution achievable for EWPC structure determination comes as the convergence angle approaches a value where the disks in the NBED pattern would overlap. At this point the probe envelope in the EWPC pattern shrinks so that the first order lattice spacing peaks are lost. EWPC lattice measurement may be done using a FWHM probe diameter just larger than the unit cell. For example, **Figure 4**(c,d) show NBED and EWPC patterns taken with a 5 mrad convergence angle, giving a probe with a 0.4 nm diffraction-limited full-width half maximum (FWHM), just larger than the 0.315 nm unit cell of $MoS_2$. The first-order spots are present and viable to fit, although they are significantly attenuated because their corresponding 0.315 nm spacing is larger than the half-width half maximum (HWHM) probe radius of 0.2 nm. Near-unit-cell spatial resolution is thus possible, although this attenuation will reduce the precision of lattice structure measurements and require a larger electron dose for high-precision measurements. With a HWHM probe radius equal to the unit cell size, giving 2-unit-cell resolution, or larger, the attenuation and precision loss will be relatively minor.

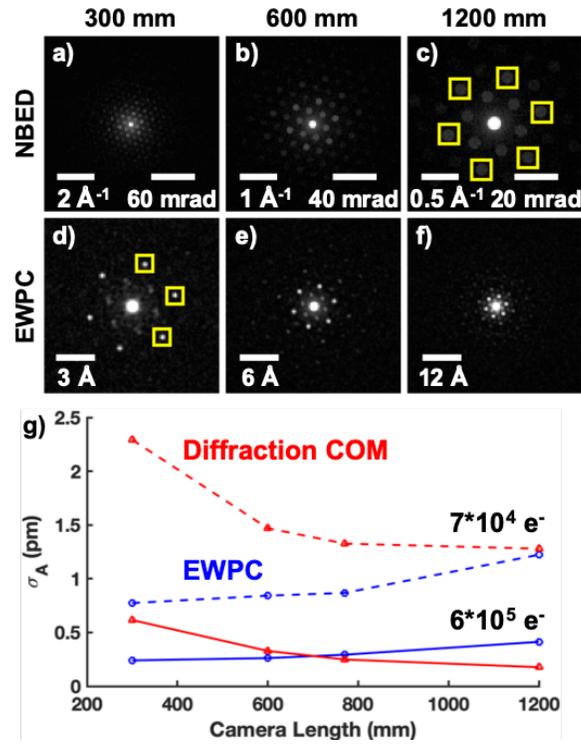

**Figure 5:** Comparison of diffraction patterns (a-c) and corresponding EWPC patterns (d-f) of monolayer $MoS_2$ at different camera lengths for a 2.5 mrad convergence angle. Camera lengths are 300 mm for (a,d), 600 mm for (b,e) and 1200 mm for (c,f). (g) Comparison of standard deviation of lattice parameter $\sigma_A$ calculated from the center of mass (COM) of diffraction spots (red) and from EWPC peak fits (blue) for different camera lengths. Dotted lines show calculation for lower-dose patterns with $\sim 7*10^4$ eletrons per pattern and solid lines are for higher-dose patterns with $6*10^5$ electrons per pattern.

Another important experimental consideration is the camera length used for NBED pattern acquisition. Changing the camera length will scale the size of the NBED pattern on the detector and the calculated EWPC pattern, as illustrated in **Figure 5** for monolayer $MoS_2$. The EWPC pattern scales inversely to the NBED pattern, with the EWPC pattern expanding as the NBED pattern shrinks. EWPC patterns using relatively small camera lengths therefore may be useful for resolving very similar lattice spacings.

The precision of lattice measurements from either direct diffraction or EWPC patterns is also affected by the choice of camera length. **Figure 5**(g) illustrates the trend in precision by plotting the standard deviation of the average lattice parameter $\sigma_A$ calculated using a center of mass (COM) approach and the EWPC. In the direct diffraction approach, which is similar to that used by Han et al., [11] we calculated the COM of the six {110} diffraction spots marked in **Figure 5**(c) and determined the three lattice vectors from the difference in the opposing spot positions. In the EWPC approach, we calculated the three lattice vectors from the positions of the three peaks marked in **Figure 5**(d) (the remaining three in the set are numerically identical). At each spatial position, the overall lattice constant $A$ was calculated as the average of the 3 lattice vector magnitudes. Diffraction patterns were collected from ~3000 distinct spatial points across a single-grain region of monolayer $MoS_2$ each separated by a ~9 nm step size. The same number of patterns were collected from the same sample region for four different camera lengths. To allow a comparison at both relatively low and high dose conditions the calculation was repeated using the same patterns after 3X3 binning in real space to increase the total electron counts from ~$7*10^4$ electrons per pattern to ~$6*10^5$ electrons per pattern (corresponding to beam currents of ~10 pA and ~90 pA for 1 ms acquisitions). The sample may have some intrinsic variations from point defects or buckling of the suspended sample, but these result in less than 1 pm spread and the same datasets were used to calculate both COM and EWPC lattice parameters.

Figure 5g shows the EWPC fitting has better precision at lower camera lengths, which allow the inclusion of more diffraction spots in the EWPC to improve the peak signal-to-noise ratio (SNR). The diffraction COM shows better precision at higher camera lengths where the diffracted disks are spread over more pixels. The precision of the diffraction COM technique degrades markedly at the lowest 300 mm camera length, where the diffracted disks are only ~2 pixels across. With the higher dose, both approaches provide precision well below 1 pm and vary slightly with camera length. At the lower dose, EWPC fitting outperforms diffraction COM, achieving sub-pm precision at lower camera lengths. This difference is likely due to the inclusion of electrons from all diffraction spots in the calculation of the first-order EWPC peaks. Even though monolayer materials such as $MoS_2$ represent a best-case scenario for the COM measurements because they do not display strong diffraction artifacts, EWPC fitting still outperforms COM at low beam current or at small camera lengths.

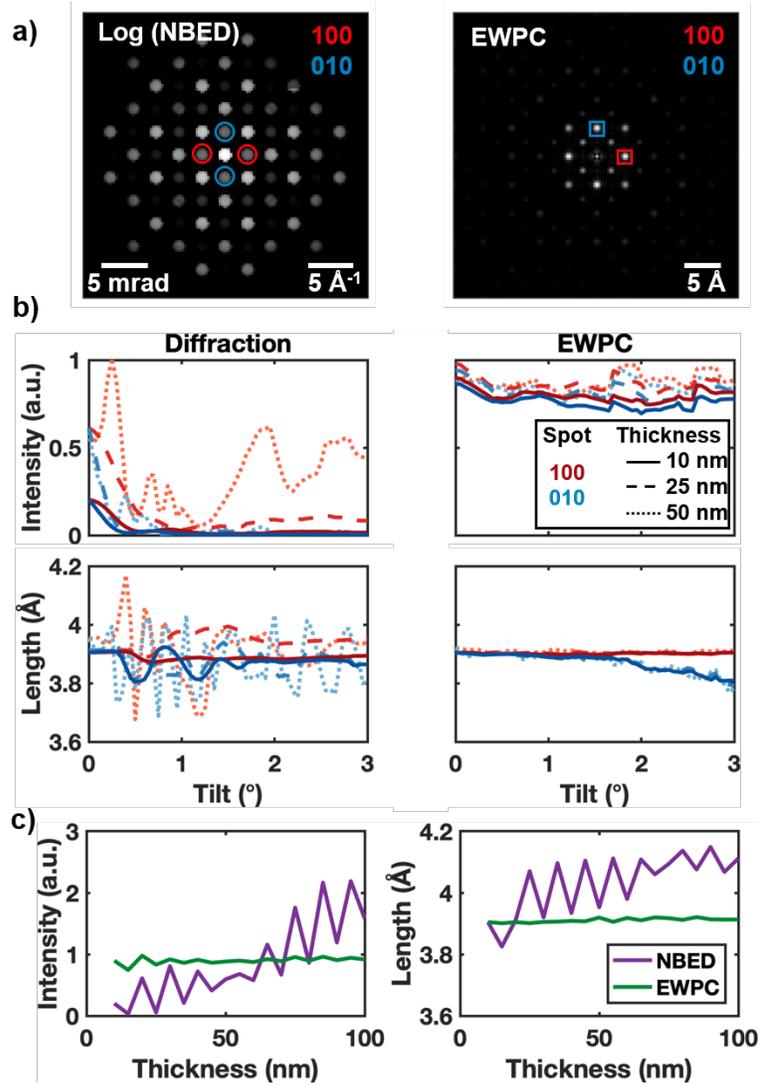

**Figure 6:** Bloch-wave simulations of the NBED and the corresponding EWPC pattern (a) for SrTiO$_3$ shown for 0° tilt and 10 nm thickness. b) Variation in the intensity (top) and lattice parameter (bottom) calculated from the selected NBED and EWPC spots shown in (a) for different sample tilts and thicknesses. c) Variation in the intensity and lattice parameter vs thickness, measured from the selected NBED and EWPC spots for an untilted sample. The EWPC measurement of lattice constant has relatively little thickness dependence.

*4.2    Dynamical Effects of Thickness and Tilt on EWPC Patterns*

To benchmark the performance of the EWPC method while accounting for dynamical diffraction effects, we performed a Bloch-wave simulation of NBED diffraction from a SrTiO$_3$ specimen over a range of tilts and thicknesses (Figure 6). Here the NBED is simulated down the [001] zone axis and tilted around the [100] zone. In Figure 6b we compare the intensity of diffraction and EWPC spots as well as measurements of the lattice parameter made from the diffraction pattern using the spot COM (left) and the EWPC pattern (right). Because of dynamical diffraction effects, the intensity of the diffracted discs oscillates strongly with tilt and thickness. Intensity variations within the diffracted disks cause the lattice parameter measured from the diffraction COM to vary with both sample tilt and thickness, with inaccuracies of ~10 pm or more. These oscillations appear for both lattice parameter directions and become more severe for greater sample thicknesses. Furthermore, the strong intensity loss in the diffraction spots under some conditions may lead to a poor signal-to-noise ratio in practical experiments.

In contrast, the EWPC spots are more stable, showing smaller variations in intensity and measured lattice parameter. The stability of the EWPC pattern despite instability in the diffraction pattern results from the logarithm and Fourier transform in its calculation, which averages out dynamical variations across many diffraction spots. The lattice parameter measured from the EWPC shows little variation (~1 pm) with thickness (Figure 6c). The intensity of the EWPC spots also show little thickness variation as the cepstral transform transfers intensities to the lower spatial frequencies (i.e. central peak) through the envelope function. The spots outside the central peak are best thought of as a type of normalized pair-correlation function.

In the [100] spacing, which is oriented parallel to the tilt axis, there is minimal variation in the lattice parameter with tilt up to 3° (Figure 6b). The [010] spacing shows variation of only ~1 pm below ~1.5° (26 mrad), but at higher tilts the measurement becomes systematically lower. The error appears to approximately follow the cosine of the tilt angle, but with a greater magnitude than expected from geometric foreshortening alone. At these relatively large tilts some of the assumptions for both our method and the Bloch wave simulation may break down, so the source of this systematic error is currently unknown, but may be explained and corrected in future investigations. Despite this, the EWPC provides significantly improved robustness in measurement, especially for mistilts less than ~1.5°. For samples oriented far from zone axis, use of the Lagrange strain can help avoid potential inaccuracies in the absolute lattice parameter measured with the EWPC method.

### 4.3 *Lattice Mapping of Ferroelectric Domains in PbTiO$_3$*

Ferroelectric materials often have atomic displacements in the crystal that drive the ferroelectric polarization, and require microscopic techniques such as atomic resolution imaging to determine local polarization. Additionally, many ferroelectric systems experience ferroelastic distortions of their unit cell on the order of a few percent of the lattice parameters and rotations of a few degrees, making strain mapping useful as a proxy for ferroelectric domains. Ferroelectric domains span wide length scales from nanometers to microns and can have atomically abrupt domain walls with emergent properties that may not be present in the bulk. While studying the domain and domain wall structure by STEM is an important step in understanding these materials, ferroelastic strain and other complexity in the specimen may introduce challenges such as varying crystallographic tilt which requires more robust characterization methods than direct lattice imaging with atomic resolution STEM. Lattice mapping with nanobeam electron diffraction and the EWPC transform combines the potential for high spatial resolution, large fields of view, and robust measurement to address these challenges.

Here we demonstrate this approach for epitaxially-grown PbTiO$_3$ films. PbTiO$_3$ has a tetragonal unit cell with its long axis in the direction of the ferroelectric polarization. Here we will examine two types of ferroelectric-ferroelastic domain arrangements: periodic a- and c-domains, alternating lattice elongation along the in-plane and out-of-plane directions; and periodic a$_1$- and a$_2$-domains alternating lattice elongation along [100] and [010] in-plane directions. The a/c and the a$_1$/a$_2$ domain configurations are formed as a mechanism to accommodate the epitaxial strain when PbTiO$_3$ is grown on lattice-mismatched DyScO$_3$ and GdScO$_3$ substrates, respectively. Details on this material system and the domain structure can be found in a separate publication.[33]

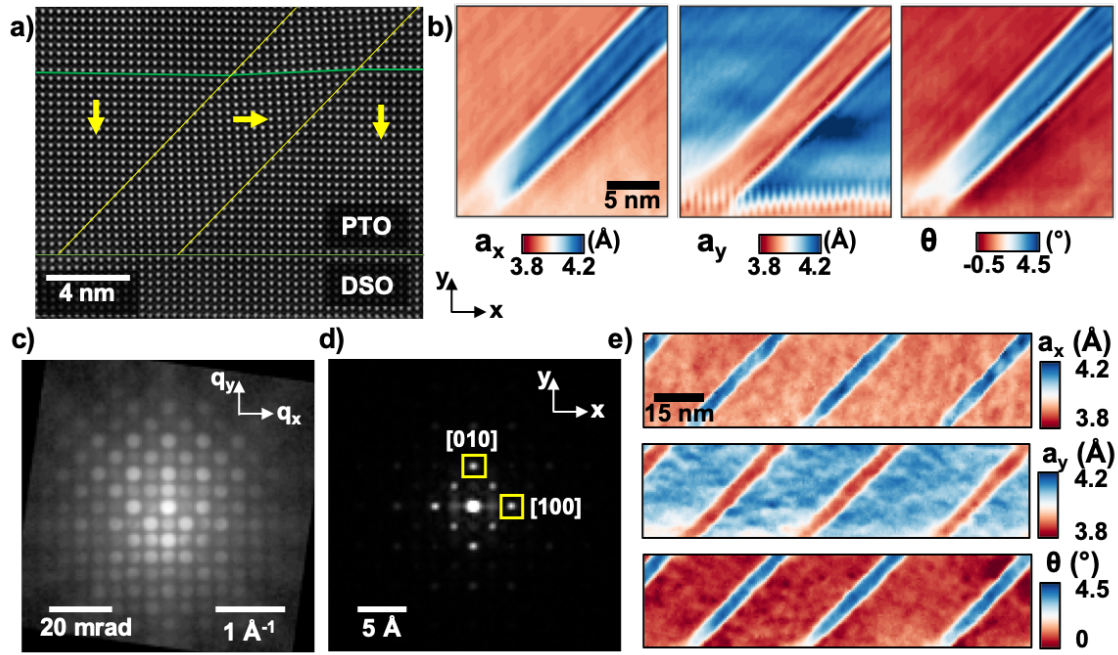

**Figure 7:** Lattice parameter mapping of ferroelastically strained PbTiO$_3$ by AC-STEM and EWPC. a) HAADF-STEM atomic resolution image of the PbTiO$_3$ (PTO) on a DyScO$_3$ (DSO) substrate. By inspecting the Ti displacement, the polarization direction can be found. At ferroelectric domains, a rotation of the lattice is observed. b) lattice parameters measured from fitting the atomic positions in the in-plane ($a_x$) direction, out-of-plane direction ($a_y$), and the rotation of the unit cell, $\theta$. c) the NBED pattern of a single domain and d) the corresponding EWPC pattern. By mapping the spots indicated in d) the lattice parameters and rotation can be measured as shown in e).

**Figure 7** presents lattice mapping of this domain structure for PbTiO$_3$ on a DyScO$_3$ substrate using the EWPC approach and aberration-corrected STEM (AC-STEM) imaging and atom-fitting for validation. **Figure 7**(a) shows an AC-STEM image of the PbTiO$_3$/ DyScO$_3$ domain structure with wide c-domains at the left and right where the long axis of the unit cell is oriented out of the plane of the film (in the "y" direction). These are separated by a narrower a-domain with the polarization oriented in the plane of the film (in the "x" direction in the image plane) that cuts diagonally across the field of view. **Figure 7**(b) shows quantitative lattice maps calculated from atom-fitting, including the lattice parameters $a_x$ and $a_y$ and the lattice rotation $\theta$. These maps confirm that the structure consists of domains where the tetragonal unit cell's long

4.156 Å axis and short 3.902 Å axis alternate directions together with a 3° rotation between the domains.

**Figure** 7(c) and (d) show NBED and EWPC patterns for a PbTiO$_3$ c-domain with the EWPC peaks used for fitting marked with yellow boxes. **Figure** 7(e) shows maps of the lattice parameters and lattice rotation calculated by tracking the EWPC peaks, which are both qualitatively and quantitatively consistent with the lattice maps from atomic resolution imaging. Comparing the lattice parameters in the c-domain, we measure 3.90 ± 0.02 Å in-plane and 4.09 ± 0.05 Å out-of-plane for atomic resolution imaging, and 3.90 ± 0.01 Å in-plane and 4.10 ± 0.02 Å out-of-plane for EWPC mapping. Mapping using the EWPC technique provides the additional advantages of allowing large fields of view with a 3-4 orders of magnitude lower beam dose.

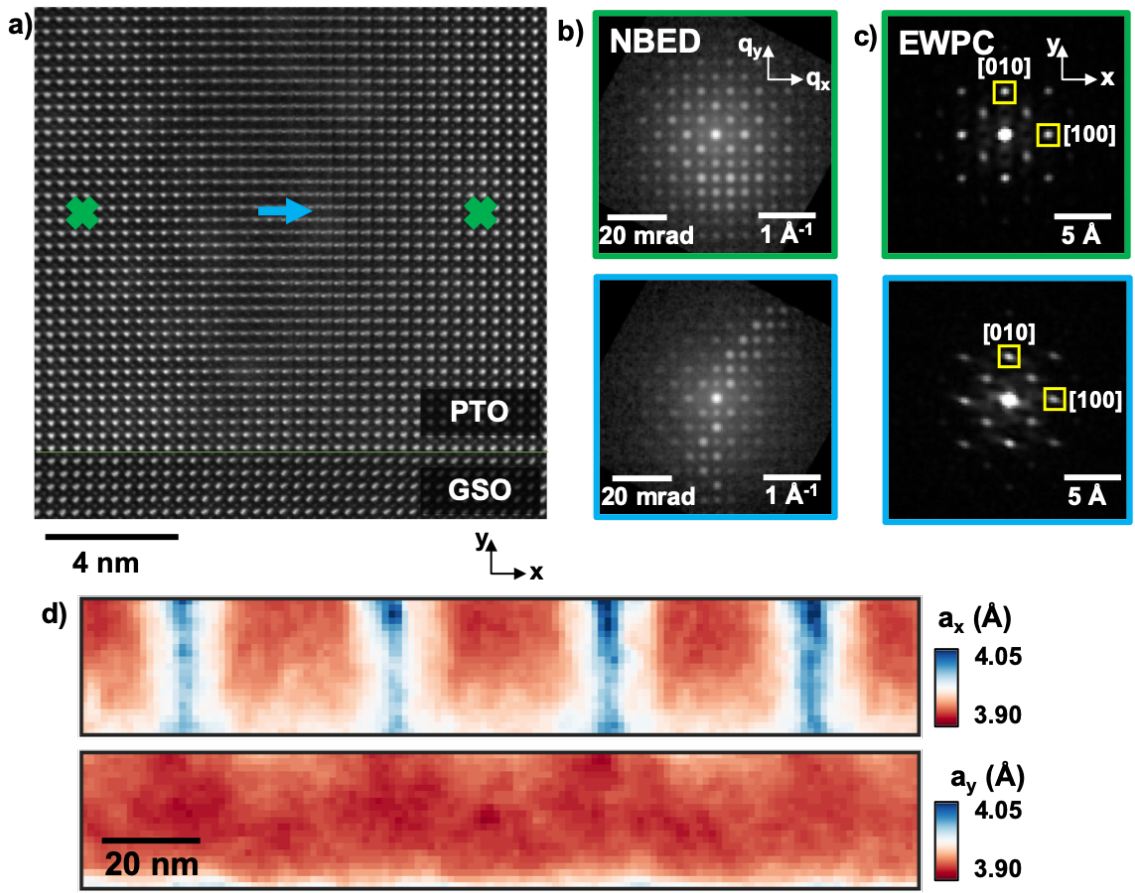

**Figure 8:** Mapping the lattice parameter of on and off-axis regions of PbTiO$_3$ that occur in ferroelectric domain configurations of PbTiO$_3$ / GdScO$_3$ (PTO/GSO). a) Atomic resolution STEM shows that at the domain marked in blue, the crystal tilts off axis and atomic resolution is lost. b) NBED and c) EWPC patterns from the regions of the crystal corresponding to ones like marked in green and blue in a). NBED shows a clear tilting off axis in the blue region, but the EWPC spots remain clear. d) EWPC is still able to map the in-plane ($a_x$) and out-of-plane ($a_y$) lattice parameters.

Depending on the film-substrate lattice mismatch, a domain structure may form in which some domains tilt in the plane of the film, such that the domain structure cannot be visualized with both domains simultaneously on axis. **Figure 8** illustrates this situation with a PbTiO$_3$ film on a GdScO$_3$ substrate. The lattice structure of one domain cannot be resolved in atomic resolution imaging (**Figure 8**(a)), and NBED patterns (**Figure 8**(b)) show that this domain has a relative tilt of ~60 mrad or ~3.4°. This presents a major challenge for characterization of this domain structure by AC-STEM. However, in the EWPC patterns for the two domains (**Figure 8**(c)) the atomic-spacing peaks remain quite sharp even for the off-axis domain. While some blurring of one set of peaks is present, which may contribute to to the excessive foreshortening error shown in Figure 6, it does not prevent domain mapping from the EWPC patterns, shown in **Figure 8**(d). These maps reveal the domain structure to be an $a_1$-$a_2$ type, with wide domains oriented with their long axis in the film plane along the optic axis and narrow domains oriented in the film and image planes ("x" direction).

*4.4   Strain Mapping of Core-Shell Pt-Co Catalyst Nanoparticles*

Another application space where high-resolution strain characterization can be highly valuable is catalytic nanoparticles. Application of strain to a catalytic surface to alter chemical binding strengths is an important strategy to enhance the catalytic activity. [34–36] Nanoparticles are commonly used as catalytic materials because of their high surface area to volume ratios and strain engineering can be achieved through core-shell nanoparticle structures. Direct lattice imaging can be used effectively to characterize lattice structure and strain for some nanoparticles, such as strongly faceted, unsupported particles that can easily be aligned and imaged on zone axis. [37–39] Supported catalysts, which are used in applications such as chemical processing and fuel cells, tend to have catalyst particles at random orientations, especially for non-crystalline supports such as carbon blacks. Furthermore, these materials are frequently heterogeneous, and high throughput characterization is needed to ensure statistically representative measurements. These concerns make direct imaging methods poorly suited for supported catalyst nanoparticles while NBED-EWPC strain mapping excels.

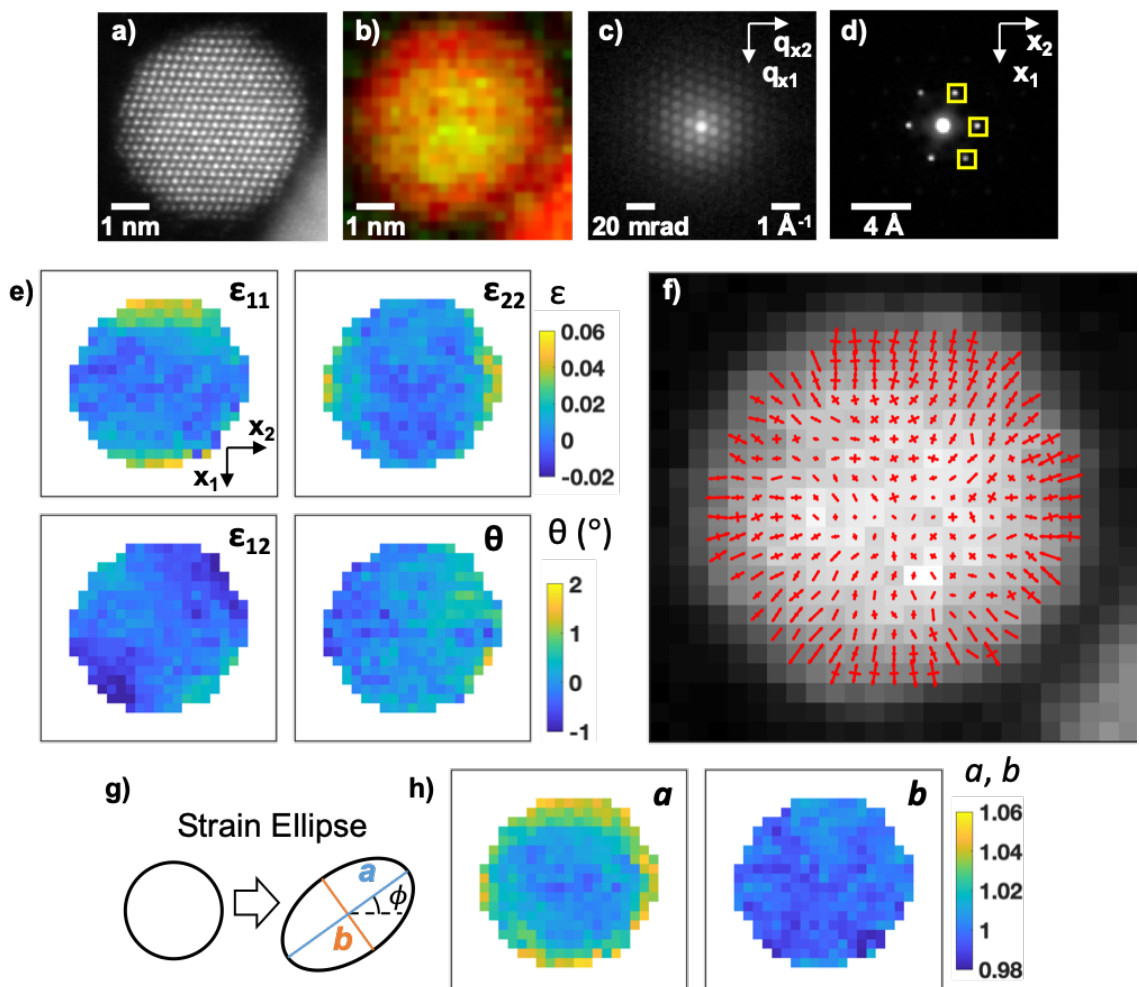

**Figure 9:** Strain mapping of of a core-shell Pt-Co fuel cell catalyst nanoparticle. (a) Atomic resolution STEM image of the particle oriented on the [110] zone axis. (b) EELS composition map showing Pt in red and Co in green. Microprobe STEM conditions were used for NBED mapping of the same particle. (c) Averaged NBED pattern and (d) corresponding EWPC pattern from the center of the particle. EWPC spots circled in red in (d) were used for strain mapping. (e) Maps of strain tensor elements for Lagrange strain referenced to the particle center. (f) Visualization of principal strain vectors as red arrows for both the principal and orthogonal secondary strain directions. (g) Schematic of the strain ellipse construction, with semi-major axis $a$, semi-minor axis $b$, and angle of principal strain $\phi$. (h) Visualizations of the strain ellipse axes $a$ and $b$.

To demonstrate the effectiveness of EWPC strain mapping for catalytic nanoparticles, we will examine the strain distribution in Pt-Co/C fuel cell catalyst nanoparticles. Pt-alloy fuel cell catalyst nanoparticles are generally prepared with a thin Pt shell which experiences a compressive strain from the smaller lattice constant of the alloy core. [40–43] This compressive strain is intended to decrease the oxygen binding strength on the surface to enhance the oxygen reduction activity. One challenge to this approach is that the catalyst particles degrade and lose catalytic activity during use because of electrochemical cycling and the harsh acidic environment in the fuel cell, which lead to growth of the Pt shells to thicknesses ranging from one to several nanometers. [30,44] Growth in the shell thickness is expected to lead to relaxation of the surface strain, resulting in loss of catalytic activity, but the details of this process are poorly understood because of the lack of suitable strain characterization.

To simulate a lifetime of fuel cell use, the Pt-Co/C catalyst examined here has been subjected to an electrochemical stability test, as described in a separate publication. [30] **Figure 9**(a) shows a ~5 nm diameter nanoparticle oriented on the [110] axis. An electron energy loss spectroscopy (EELS) map (**Figure 9**b) of the Co $L_{2,3}$ and Pt $M_{4,5}$ edges shows that a ~1 nm thick Pt shell is present around the Pt-Co core. Compressive strain in the Pt shell is visible in the AC-STEM image (**Figure 9**(a)) as a slight inward bowing of the near-surface lattice planes. A NBED map of this particle was acquired using a 1.75 mrad convergence angle giving a 0.7 nm diffraction-limited resolution. An averaged NBED pattern and corresponding EWPC pattern are shown in **Figure 9**(c) and (d), respectively. Three strong peaks in the EWPC pattern, marked in **Figure 9**(d), were tracked for strain mapping. These peaks represent interatomic spacings in the projected lattice, corresponding to the $\left[\frac{\bar{1}}{4}, \frac{\bar{1}}{4}, \frac{1}{2}\right]$, $\left[\frac{1}{4}, \frac{\bar{1}}{4}, \frac{1}{2}\right]$, and $\left[\frac{1}{2}, \frac{\bar{1}}{2}, 0\right]$ vectors in the real face-centered cubic lattice.

The Lagrange strain was calculated using the center of the particle as a reference, and the resulting strain tensor components and lattice rotation are shown in **Figure 9**(e). The primary effect visible is the radially outward Poisson expansion resulting from the tangential compression of the Pt shell. In rectangular coordinates, this manifests as high $\varepsilon_{11}$ at the top and bottom, high $\varepsilon_{22}$ at left and right, and $\varepsilon_{12}$ that is high at top-left and bottom right and low at top-right and bottom left. The changing strain direction around the shell can make these tensor components difficult to interpret. An alternative is to visualize principal strain vectors, as shown in **Figure 9**(f) which can more directly show that the strain is primarily a radially outward expansion in the shell. The principal strain can also be presented using the strain ellipse construction, which is formed by applying the strain transformation to a unit circle, as illustrated in **Figure 9**(g). The semi-major and semi-minor axes of the strain ellipse, *a* and *b*, (**Figure 9**(h)) indicate the magnitudes of the two principal strain vectors. For this approximately spherical core-shell geometry, the semi-major axis is approximately the radial strain, and the semi-minor axis is approximately the tangential strain. By adjusting to the varying strain direction, the strain ellipse axes produce a visualization of the particle's core-shell structure that is simpler to interpret.

The on-axis particle shown in **Figure 9** could in principal have its strain mapped from direct imaging, as previously reported for an unsupported Pt nanoparticle. [37] Reaching the required precision on the order of a few picometers using AC-STEM imaging requires both high SNR and correction of STEM imaging distortions, typically by image stack acquisition and registration in post-processing. These requirements impose high dose requirements – on the order of $10^8 - 10^9$ e-/nm$^2$.[37] High beam dose can be problematic as particles may rotate from damage to the support or momentum transferred from incident electrons. NBED-EWPC strain mapping is dramatically more dose efficient, with a dose of around $10^6$ e-/nm$^2$ used here. NBED-EWPC strain mapping also compares favorably on dose to EELS as a method to visualize core-shell structure, as the EELS map in **Figure 9**(b) required $10^9$ e-/nm$^2$.

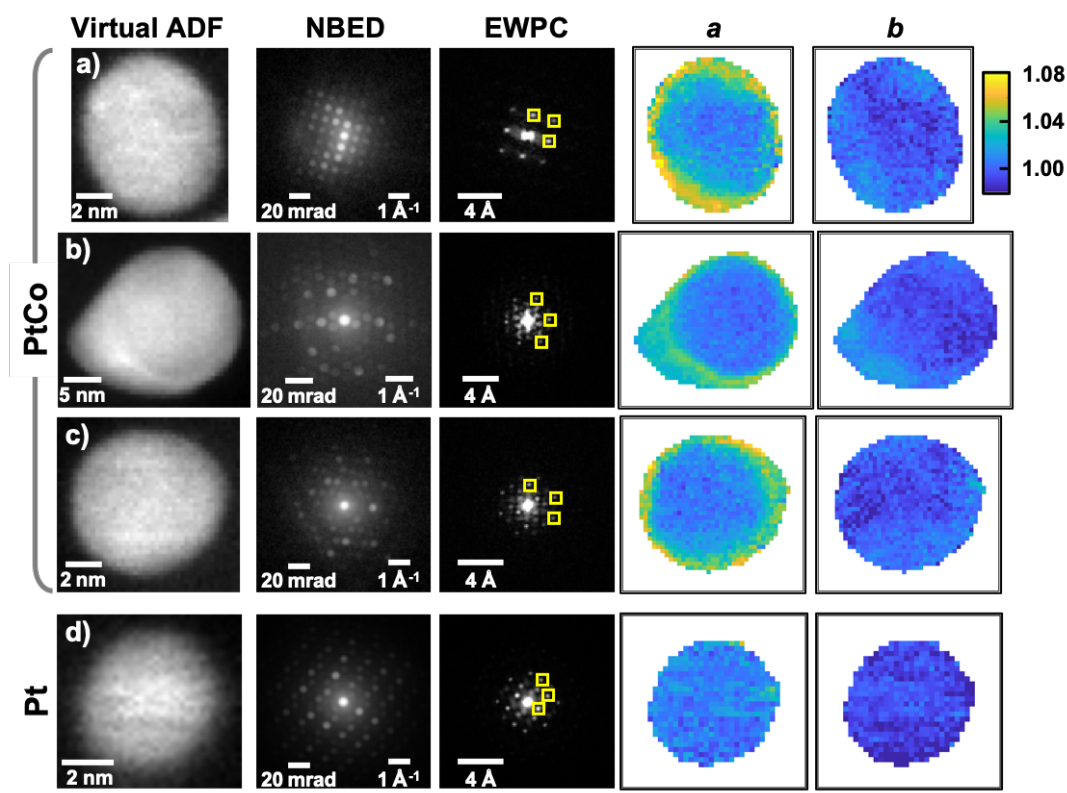

**Figure 10:** Strain mapping of three off-axis Pt-Co and Pt nanoparticles, including, from left to right, a virtual ADF image of the particle, an averaged NBED pattern from the particle center, the corresponding EWPC pattern with spots used for mapping circled in red, and the Lagrange strain referenced to the particle center visualized through the semi-major axis *a* and semi-minor axis *b* of the strain ellipse. The particles in (a-c) are Pt-Co alloys with a pure Pt shell that shows strain, while the particle in (d) is pure Pt and shows no core-shell strain. The particle shown in (a) is oriented near the [200] zone axis, while the particles in (b-d) are far from any major zone axis.

EWPC strain mapping can also overcome the major challenges of random particle orientations and varying mistilt in strained particles. Given the mechanical imprecision of TEM stages, it is generally impractical to orient any given nanoparticle onto zone axis. On-axis particles such as the one in **Figure 9** typically must be sought out and relying on these for strain mapping will compromise throughput and statistical representativeness. **Figure 10** shows four examples of particles with varying orientations. Some particles are oriented near a major zone axis, like in **Figure 10** (a), which is a couple degrees from the 200 orientation. Similar to the $PbTiO_3/GdScO_3$ example in **Figure 8**, these particles will have the same peaks visible as they would if oriented precisely on a nearby major axis. The peaks show some mild blurring, but are easily sharp enough for mapping. More commonly, particles selected at random are not close to any major zone axis, as in **Figure 10**(b) and (c). However, these patterns nearly all have sharp peaks suitable for tracking and calculation of the Lagrange strain. Varying mistilt caused by lattice distortions also generally do not cause EWPC peaks to disappear, in contrast to diffraction spots, making both random particle orientations and varying internal mistilts not significant obstacles to producing strain maps. For the Pt-Co alloy particles in **Figure 10**(a-c), the strain maps produced by this method reveal the strain profiles formed in the Pt shells strained by the underlying Pt-Co cores. In contrast, the pure Pt particle in **Figure 10**(d) shows no core-shell strain profile.

These methods are also suitable for high throughput characterization of nanoparticle strain. A typical 256X256 NBED map takes ~2 minutes to acquire and can contain several particles suitable for strain mapping. The example maps shown here typically have ~40X40 real space pixels for a single particle map. With the high-end desktop computer used for this study, less than 30s of computation was needed to produce a strain map of a nanoparticle with the off-particle pixels excluded. With only a few minutes needed per map between acquisition and computation, the throughput of this technique is at least as high as EELS mapping of the core-shell structure for the same system, which has previously been used for a statistical study including hundreds of nanoparticle composition maps.[44] Strain mapping with the EWPC transform and EMPAD detector is a practical method to collect and analyze tens or hundreds of particle maps as needed for statistics.

## 5   Summary and Outlook

In this manuscript we have demonstrated the EWPC transform for NBED to enable robust mapping of strain and lattice structure in complex specimens. We presented a simplified analytical model to aid in interpretation of transformed patterns. This model indicates that the EWPC transform allows additive separation of the effects of tilt and thickness from the lattice information contained in periodic diffraction spots. The resulting peaks correspond to real-space inter-atomic spacings, similar to a Patterson pair-correlation function. The EWPC peaks are attenuated by a function corresponding to the real-space probe, and first-order inter-atomic spacings can be measured for probe sizes just above the unit cell size. This allows the method to attain near-unit-cell resolution, which is sub-nm for many materials. Fitting of the EWPC peaks can provide measurements of the lattice parameters with sub-pm precision. Precision is optimum for NBED patterns acquired with low camera lengths to allow inclusion of many diffraction spots. Measurement of lattice structure from EWPC peaks is also highly dose efficient, using ~1000X less dose than comparable atomic-resolution STEM measurements.

We have demonstrated the practical use of the EWPC transform for mapping ferroelectric-ferroelastic domains in strained $PbTiO_3$ films and compressively strained Pt shells in core-shell Pt-Co nanoparticle fuel cell catalysts. The distortions in the $PbTiO_3$ films measured with the EWPC technique were consistent with measurements by atomic-resolution STEM atom-fitting, while providing a larger field of view and lower beam dose. Use of the EWPC technique provided the additional benefit of allowing quantitative lattice mapping in domains tilted ~3-4° off zone-axis, which could not be resolved in atomic-resolution STEM. For the core-shell Pt-Co nanoparticles, the EWPC technique allowed sub-nm resolution mapping of strain in Pt shells thinner than a few nanometers. The technique was further demonstrated to be capable of mapping strain in off-axis particles and providing high-throughput analysis for statistically representative study of heterogeneous nanoparticle samples.

The EWPC transform for NBED mapping has the potential to be a useful tool in a wide range of applications. The sharp interatomic spacing peaks make precise fitting for lattice structure extraction numerically convenient, which may be beneficial even when precession is used. The EWPC properties allowing separation of useful lattice information from tilt and thickness effects make the technique robust for application to complex and challenging specimens. The EWPC peaks maintain stable intensity to allow precise mapping even with varying mistilt, and the presence of sharp peaks when far from major zone axis enable strain mapping in randomly oriented crystallites.

Essential functions and example MATLAB codes for power-cepstral STEM (PC-STEM) analysis are available on Github at https://github.com/muller-group-cornell/PC-STEM.

# 6 Appendix

## 6.1 Comparison of the EWPC to the Patterson Function

The exit wave power cepstrum (EWPC) is closely related to the Patterson function, which used for atomic structure determination from diffraction patterns. The Patterson function $PF(\boldsymbol{x})$ can be calculated from a diffraction pattern $I(\boldsymbol{q})$ as the Fourier transform of the diffraction pattern:
$$\mathrm{PF}(\boldsymbol{x}) = |\mathcal{F}(I(\boldsymbol{q}))|. \tag{17}$$
The key difference between the Patterson function and the EWPC is the addition of the logarithm in the EWPC:
$$\mathrm{EWPC} = \left|\mathcal{F}\left(\ln(I(\boldsymbol{q}))\right)\right|. \tag{18}$$
This provides the useful separation properties of the EWPC transform described in Section 2 of the main text, which give it advantages over the Patterson function for nanobeam electron diffraction.

To contrast with the Patterson function, we will attempt to follow the logic of Section 2.2 for the interpretation of Patterson functions calculated from nanobeam electron diffraction patterns. Following Eq. 9, we can approximate the diffraction pattern in terms of the reciprocal space probe function $\Phi_P(\boldsymbol{q})$, the projected potential $\mathcal{V}_0(\boldsymbol{q})$, and the Ewald sphere envelope function $E(\boldsymbol{q})$:
$$I(\boldsymbol{q}) \approx |E(\boldsymbol{q})|^2 \cdot |\Phi_P(\boldsymbol{q}) * \mathcal{V}_0(\boldsymbol{q})|^2. \tag{19}$$
Using this model of the diffraction pattern and following Eq. 10,
$$\begin{aligned}\mathrm{PF}(\mathbf{x}) &\approx |\mathcal{F}(|E(\boldsymbol{q})|^2 \cdot |\Phi_P(\boldsymbol{q}) * \mathcal{V}_0(\boldsymbol{q})|^2)| \\ &= |\mathcal{F}(|E(\boldsymbol{q})|^2) * \mathcal{F}(|\Phi_P(\boldsymbol{q}) * \mathcal{V}_0(\boldsymbol{q})|^2)|,\end{aligned} \tag{20}$$
the envelope term does not additively separate, as in the EWPC case, but remains convolved into lattice structure and probe terms. Using the probe approximation to simplify as in Eq. 11, the probe function may be extracted from the inner convolution at the right:
$$\begin{aligned}\mathrm{PF}(\mathbf{x}) &= |\mathcal{F}(|E(\boldsymbol{q})|^2) * \mathcal{F}(\Phi_P(\boldsymbol{q}) * |\mathcal{V}_0(\boldsymbol{q})|^2)| \\ &= \left|\mathcal{F}(|E(\boldsymbol{q})|^2) * \left(\phi_P(\boldsymbol{x}) \cdot \mathcal{F}(|\mathcal{V}_0(\boldsymbol{q})|^2)\right)\right|.\end{aligned} \tag{21}$$
The probe function thus behaves similarly for the Patterson function and EWPC, attenuating at large real-space spacings. However, without the separation properties of the EWPC, the envelope function $E(\boldsymbol{q})$ is convolved into the useful lattice information in the Patterson function, as illustrated in Figure 11. This makes the interatomic spacing peaks broader, and more difficult to distinguish (Figure 11c) in comparison to an equivalent EWPC pattern (Figure 11b). This is especially troublesome for a tilted (or thick) specimen, such as shown at right in Figure 11, where the envelope function is broadened, leading to a degradation of the interatomic spacing peaks.

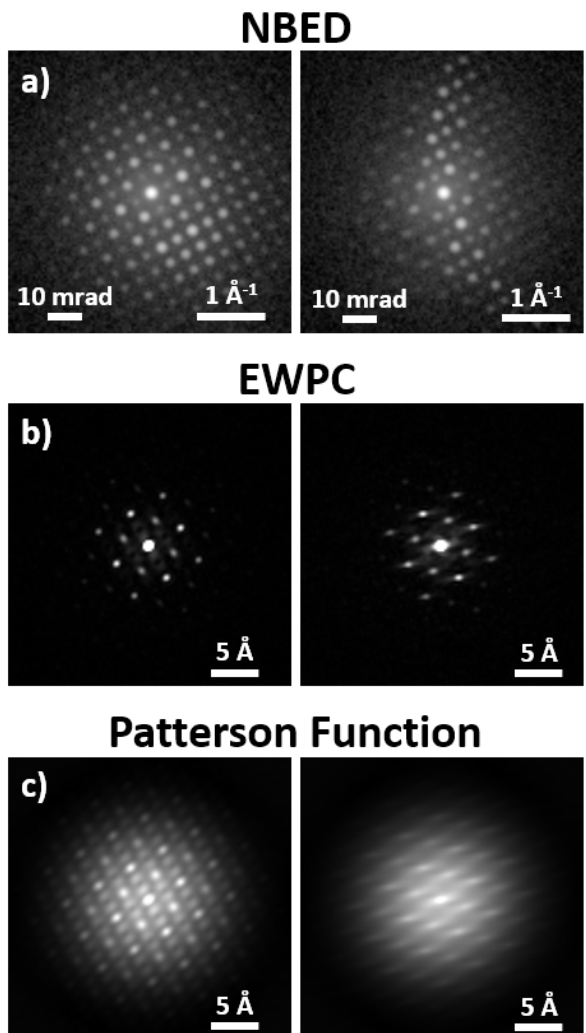

**Figure 11:** Illustration of NBED patterns (a) from on axis (left) and off-axis (right) domains of PbTiO$_3$ for comparison of EWPC (b) and Patterson function (c) transforms calculated from the NBED patterns.

## 7 Acknowledgements

This work was supported by the U.S. Department of Energy, Office of Energy Efficiency and Renewable Energy under grant DE-EE0007271, and through the Center for Alkaline-based Energy Solutions, an Energy Frontier Research Center funded by the US Department of Energy, Office of Science, Basic Energy Sciences Award DE-SC0019445. Elliot Padgett acknowledges support from an NSF Graduate Research Fellowship (DGE-1650441). Paul Cueva acknowledges funding from the Center for Bright Beams (NSF award PHY-1549132). Megan Holtz, Eric Langenberg, and Darrell Schlom acknowledge support from the US Department of Energy, Office of Basic Energy Sciences, Division of Materials Sciences and Engineering, under Award No. DE-SC0002334. Eric Langenberg acknowledges funding from the EU's Horizon 2020 program through the Marie Skłodowska-Curie Actions: Individual Fellowship-Global Fellowship

(ref. MSCA-IF-GF-708129). Yu-Tsun Shao and David Muller acknowledge support from the AFOSR Hybrid Materials MURI, under award number FA9550-18-1-0480. This work made use of electron microscopy facilities supported by the NSF MRSEC program (DMR-1719875) and an NSF MRI grant (DMR-1429155). The authors thank John Grazul, Malcolm Thomas, and Mariena Silvestry Ramos for assistance with electron microscopy facilities and sample preparation, as well as Anusorn Kongkanand at General Motors for providing fuel cell catalyst samples, and Ariana Ray, Swathi Chandrika, and Don Werder for the molybdenum disulfide samples. The authors thank Michael Cao, Zhen Chen, Yimo Han, and Kayla Nguyen for useful discussions and assistance with the EMPAD.